\documentclass[reprint,aps,prl,notitlepage,longbibliography]{revtex4-2}

\usepackage{graphicx}
\usepackage{float}
\usepackage{hyperref}

\begin{document}
\textbf{Comment on ``Origin of correlated isolated flat bands in copper-substituted lead phosphate apatite''} by Michael W.~Swift and John L.~Lyons, Center for Computational Materials Science, US Naval Research Laboratory

Griffin~\cite{griffin_origin_2023} presented first-principles calculations using PBE+U showing that in Pb$_{10}$(PO$_4$)$_6$(OH)$_2$, copper substitution on the Pb(1) site gives rise to a pair of bands approximately 160 meV above the host material's valence-band maximum (VBM).  This two-band manifold is half occupied and very flat, resulting in a large density of states at the Fermi level.  Such an electronic structure is a characteristic of many high-$T_\mathrm{c}$ superconductors, and was suggested as a possible explanation for recent experimental reports of room-temperature superconductivity in so-called LK-99~\cite{lee_first_nodate,lee_superconductor_nodate,si_electronic_2023,cabezas-escares_theoretical_2023,kurleto_pb-apatite_2023}.

\begin{figure}
\includegraphics[width=\columnwidth]{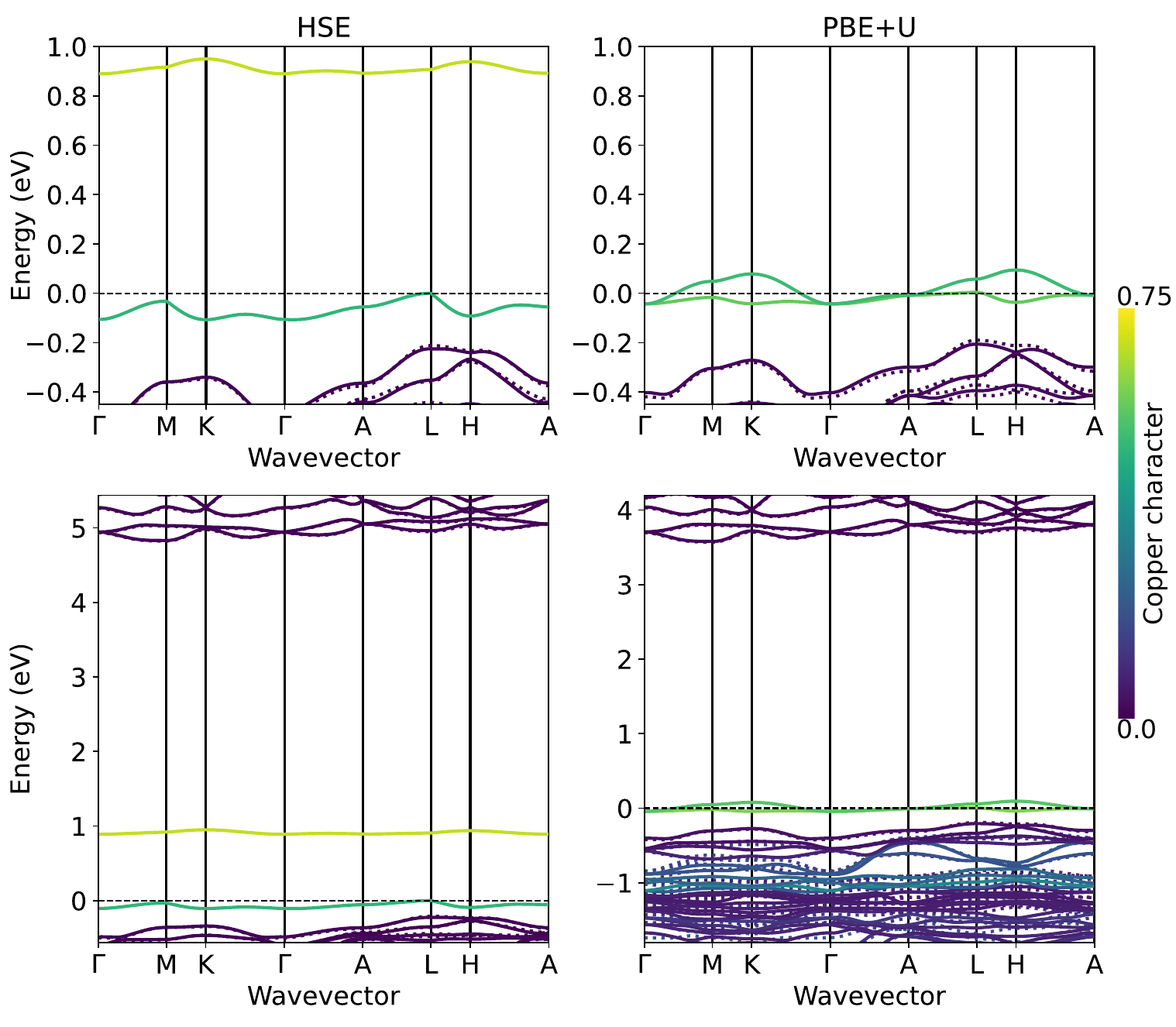}
\caption{Band structure of copper-substituted lead phosphate apatite with copper on the Pb(1) site (Cu$_\mathrm{Pb(1)}$).  Bands are color-coded according to the fraction of the charge density that projects onto the copper orbitals.  Results using HSE are on the left and PBE+U results are on the right.  The top plots show details near the Fermi level (set to zero and indicated with a dotted line).  The bottom row shows the full gap, vertically aligned based on the conduction band.  HSE predicts an insulating state: the VBM consists of a fully occupied copper state, and an unoccupied copper state lies about 0.9 eV higher in energy.  This contrasts with the half-occupied two-band manifold in PBE+U.}
\label{figure}
\end{figure}

Our calculations show that copper substituting on the Pb(1) site (which we will call Cu$_\mathrm{Pb(1)}$) does not give rise to a half-occupied flat band.  Instead, the two copper bands are split into a lower, fully occupied band (which forms the VBM) and an upper, unoccupied band (see Figure~\ref{figure}).  The difference is due to our use of the HSE hybrid functional~\cite{HSE}, which provides more accurate band-gap energies and generally provides a better description of dopants in semiconductors.  Other calculation parameters were similar, and our tests using PBE+U reproduce the results of Ref.~\citenum{griffin_origin_2023}~\footnote{Our HSE calculations used VASP, PAW pseudopotentials, a 500 eV plane-wave energy cutoff, a 2$\times$2$\times$3 k-point mesh, and standard ``HSE06'' parameters.  Relaxations achieved force convergence of 0.03 eV/\AA.  PBE+U calculations used $U=4$~eV on the copper d states. HSE-relaxed and PBE+U-relaxed structures were very similar, and HSE calculation of the PBE+U-relaxed structure yielded nearly identical electronic structure to the full HSE calculation.}.  Semilocal functionals like PBE are known to underestimate band gaps, in oxides largely by predicting oxygen 2p bands that are too high in energy.  PBE can therefore predict defect and impurity states to be resonant with the valence band when they should be localized in the band gap~\cite{janotti_hybrid_2010}. In Ref.\citenum{griffin_origin_2023} PBEsol+U was used, thus taking electron correlations into account by incorporating an onsite Hubbard interaction $U$ for the copper impurity.  While this approach can correct the band gap and defect levels in some circumstances~\cite{lany_predicting_2011}, it would provide no correction for the energy of the oxygen p states.  We find that PBE underestimates the HSE-calculated band gap of the parent compound, lead phosphate apatite, by almost 1.3 eV~\footnote{The PBE gap of Pb$_{10}$(PO$_4$)$_6$(OH)$_2$ is 3.61 eV compared to an HSE gap of 4.89 eV}.

Our results suggest that the half-occupied flat band is an artifact of the PBE overestimation of the valence band energy.  The occupied copper state (which we observe using HSE) is hybridized with oxygen, so its energy is overestimated in PBE+U along with the valence band.  This pushes the state up in the band structure and brings it close in energy to the unoccupied copper state, causing the two copper states to mix, and resulting in a half-occupied flat band.   Within HSE, the two copper states are separate.  This also explains why varying the copper $U$ value between 2 and 8 eV did not significantly shift the relative positions of the flat band and the valence band for PBE+U in Ref.~\citenum{griffin_origin_2023}: the flat-band energy is not dictated by the applied $U$, but rather by hybridization with the oxygen states that are artificially high in energy.

Griffin also pointed out that Cu$_\mathrm{Pb(2)}$, the lower-energy substitutional site, does not lead to a half-occupied flat band, but instead exhibits oxygen 2p character at the VBM and a single unoccupied copper band at mid-gap.  Our results are qualitatively similar, both in PBE+U and HSE.  We find Cu$_\mathrm{Pb(2)}$ is lower in energy than Cu$_\mathrm{Pb(1)}$ by 0.83 eV.  For Cu$_\mathrm{Pb(2)}$, we calculate the gap between the VBM and copper state to be 0.47 eV in PBE+U and 2.10 eV in HSE; PBE+U underestimates the gap between the VBM and the copper states in Cu$_\mathrm{Pb(2)}$ by 1.6 eV.  Therefore it is not surprising that PBE+U closes the 0.9 eV gap between copper states in Cu$_\mathrm{Pb(1)}$.

Our results predict that stoichiometric Pb$_9$Cu(PO$_4$)$_6$(OH)$_2$ is an insulator, not a superconductor. But, these results on their own do not rule out the possibility of superconductivity in LK-99; with degenerate doping, the Fermi level could be pushed into one of the high density-of-states copper bands.  Given the questions surrounding the quality of the initial LK-99 material,\cite{zhu2023order} careful experiments on high-quality samples will be crucial. We have furthermore shown that density functional theory calculations at the PBE+U level predict an electronic structure that is qualitatively inconsistent with results from a hybrid functional, indicating that future first-principles studies should employ advanced methods to ensure that the oxygen-derived valence band is correctly described.

\bibliography{References}

\end{document}